\documentclass[hyper,12pt,letterpaper]{JHEP}  
\usepackage{epsfig}

\newfont{\frak}{eufm10 scaled 1200}

\newfont{\Bbb}{msbm10 scaled 1200}     
\newcommand{\mathbb}[1]{\mbox{\Bbb #1}}
\DeclareSymbolFont{AMSa}{U}{msa}{m}{n}
\DeclareSymbolFont{AMSb}{U}{msb}{m}{n}
\let\Box\relax
\DeclareMathSymbol{\Box}{\mathord}{AMSa}{"03}

\def\IZ{{\mathbb Z}}
\def\IR{{\mathbb R}}

\def \eqn#1#2{\begin{equation}#2\label{#1}\end{equation}} 

\def \lh{L_{het}}
\def \gh{g_{het}}
\def \li{L_{type\,I}}

\def \lt{{L_{planck}^{ten}}}
\def \gi{g_{type\,I}}
\def \lpl{{L_{planck}^{eleven}}}

\def\hacek{\accent20}                           

\title{On the hyperbolic structure\\
 of moduli spaces with 16 SUSYs}

\author{Lubo\hacek s Motl\\
  Department of Physics and Astronomy\\
  Rutgers University, Piscataway, NJ 08855-0849\\
E-mail: \email{motl@physics.rutgers.edu}}

\author{Tom Banks\\
  Department of Physics and Astronomy\\
  Rutgers University, Piscataway, NJ 08855-0849\\
E-mail: \email{banks@physics.rutgers.edu}}

\abstract{We study the asymptotic limits of the heterotic string theories
compactified on tori. We find a bilinear form uniquely determined by
dualities which becomes Lorentzian in the case of one spacetime dimension.  
For the case of the $SO(32)$ theory, the limiting descriptions include
$SO(32)$ heterotic strings, type~I, type~IA and other T-duals, M-theory on
K3, type IIA theory on K3 and type IIB theory on K3 and possibly new
limits not understood yet.}


\received{???????? ?st, 1998}
\accepted{???????? ?th, 1998}

\preprint{\hepth{9904008}\\RU-99-14\\HEP-UK-0008}
\begin{document}


\section{Introduction}
%
In a recent collaboration with W.~Fischler \cite{bfm},
we showed that the space of asymptotic directions in the
moduli space of toroidally compactified M-theory had
a hyperbolic metric, related to the hyperbolic structure
of the $E_{10}$ duality group.  We pointed out that this
could have been anticipated from the hyperbolic nature of
metric on moduli space in low energy SUGRA, which ultimately
derives from the negative kinetic term for the conformal 
factor.

An important consequence of this claim is that there are 
asymptotic regions of the moduli space which cannot be
mapped onto either 11D SUGRA (on a large smooth manifold) 
or weakly coupled Type II string theory.  These regions
represent true singularities of M-theory at which no known
description of the theory is applicable.  Interestingly, 
the classical solutions of the theory all follow 
trajectories which interpolate between the mysterious
singular region and the regions which are amenable to a 
semiclassical
description.  This introduces a natural arrow of time into
the theory.   We suggested that moduli were the natural
semiclassical variables that define cosmological time in
M-theory and that ``the Universe began'' in the 
mysterious singular region.

We note that many of the singularities of the classical
solutions {\it can} be removed by duality transformations.
This makes the special nature of the singular region all the
more striking\footnote{For reference, we note that there are
actually two different types of singular region: neither
the exterior of the light cone in the space of asymptotic 
directions,
nor the past light cone, can be mapped into the safe domain.
Classical solutions do not visit the exterior of the
light cone.}.

In view of the connection to the properties of the low energy
SUGRA Lagrangian, we conjectured in \cite{bfm} that the
same sort of hyperbolic structure would characterize
moduli spaces of M-theory with less SUSY than the toroidal
background.  In this paper, we verify this conjecture for
11D SUGRA backgrounds of the form $K3 \times T^6$, which
is the same as
the moduli space of heterotic strings compactified
on $T^9$.  A notable difference is the absence of a completely
satisfactory description of the safe domains of asymptotic
moduli space.  This is not surprising.  The moduli space is
known to have an F-theory limit in which there is no complete
semiclassical description of the physics.  Rather, there
are different semiclassical limits valid in different regions
of a large spacetime.    

\vspace{3mm}

Another difference is the appearance of asymptotic domains
with different internal symmetry groups.  
11D SUGRA on $K3 \times T^3$
exhibits a $U(1)^{28}$ gauge group in four noncompact dimensions.
At certain singularities, this is enhanced to a nonabelian
group, but these singularities have finite codimension in the
moduli space.  Nonetheless, there are asymptotic limits
in the full moduli space 
({\it i.e.} generic asymptotic directions) in which the
full heterotic symmetry group is restored.  From the heterotic 
point of view, the singularity removing, symmetry breaking, 
parameters are Wilson lines on $T^9$.  In the infinite
(heterotic torus) volume limit, these become irrelevant.  
In this paper we will
only describe the subspace of asymptotic moduli space with
full $SO(32)$ symmetry.  We will call this the HO moduli space from now on.
 The points of the moduli space will be
parametrized by the dimensionless heterotic string coupling constant
$\gh=\exp{p_0}$ and the radii $R_i=\lh\exp{p_i}$ where $i=1,\dots 10-d$
with $d$ being the number of large spacetime dimensions and $\lh$ denoting
the heterotic string length. Throughout the paper we will neglect
factors of order one.

\vspace{3mm}

Apart from these, more or less expected, differences, our 
results are quite similar to those of \cite{bfm}.  
The modular group of the completely compactified theory
preserves a Lorentzian bilinear form with one timelike direction. 
The (more or less) well understood regimes correspond to the
future light cone of this bilinear form, while all classical
solutions interpolate between the past and future light
cones.  We interpret this as evidence for a new hyperbolic algebra
${\cal O}$, whose infinite momentum frame Galilean subalgebra is
precisely the affine algebra $\hat{o} (8,24)$ of
\cite{sentwod}-\cite{schwtwod}.
This would precisely mirror the relation between $E_{10}$ and $E_9$.
Recently, Ganor \cite{ori} has suggested the $DE_{18}$ Dynkin
diagram as the definition of the basic algebra of toroidally
compactified heterotic strings.  This is indeed a hyperbolic
algebra in the sense that it preserves a nondegenerate
bilinear form with
precisely one negative eigenvalue\footnote{Kac' definition
of a hyperbolic algebra requires it to turn into an affine or
finite dimensional algebra when one root of the Dynkin diagram 
is cut.  We believe that this is too restrictive and that the
name hyperbolic should be based solely on the signature of the
Cartan metric.  We thank O. Ganor for discussions of this
point.}.

\subsection{The bilinear form}

We adopt the
result of \cite{bfm} with a few changes in notation. First we will use
$d=11-k$ instead of $k$ because now we start in ten dimensions instead of
eleven.  The parameter that makes the parallel between toroidal M-theory
and heterotic compactifications most obvious is the
number
of large spacetime dimensions $d$. In \cite{bfm}, 
the bilinear form was 
\eqn{bilstart}{I=(\sum_{i=1}^{k}P_i)^2 + (d-2)\sum_{i=1}^k(P_i^2).}
where $P_i$ (denoted $p_i$ in \cite{bfm}) are the logarithms of the radii
in 11-dimensional Planck units.

Now let us employ the last logarithm
$P_k$ as the M-theoretical circle of a type IIA
description. For the HE theory, which can be understood as M-theory on
a line interval, we expect the same bilinear form where $P_k$ is the
logarithm of the length of the Ho\hacek rava-Witten line interval.
Now we convert (\ref{bilstart}) to the heterotic units according to the
formulae ($k-1=10-d$)
\eqn{bildva}{P_k=\frac 23 p_0,\qquad
P_i=p_i-\frac 13 p_0, \quad i=1,\dots 10-d}
where $p_0=\ln{\gh}$ and $p_i=\ln(R_i/ \lh)$ for $i=1,\dots 10-d$.
To simplify things, we use natural
logarithms instead of the logarithms with a large base $t$
like in \cite{bfm}. This corresponds to a simple rescaling of $p$'s
but the directions are finally the only thing that we study.
In obtaining (\ref{bildva}) we have used the well-known formulae $R_{11}=\lpl
\gh^{2/3}$
and
$\lpl=\gh^{1/3}\lh$. Substituing (\ref{bildva}) into (\ref{bilstart})
we obtain
\eqn{biltri}{I=(-2p_0+\sum_{i=1}^{10-d}p_i)^2
+(d-2)\sum_{i=1}^{10-d}(p_i^2).}
This bilinear form encodes the kinetic terms for the moduli in the
$E_8\times E_8$ heterotic theory (HE) in the Einstein frame for the large
coordinates.

We can see very easily that (\ref{biltri}) is conserved by T-dualities.
A simple T-duality (without Wilson lines) takes HE theory to HE theory
with $R_1$ inverted and acts on the parameters like
\eqn{tdu}{(p_0,p_1,p_2,\dots)\to(p_0-p_1,-p_1,p_2, \dots).}
The change of the coupling constant keeps the effective 9-dimensional
gravitational constant $\gh^2/R_1=\gh'^2/R'_1$ (in units of $\lh$) fixed.
In any number of dimensions (\ref{tdu}) conserves the quantity
\eqn{pten}{p_{10}=-2p_0+\sum_{i=1}^{10-d}p_i}
and therefore also the first term in (\ref{biltri}). The second term in
(\ref{biltri}) is fixed trivially since only the sign of $p_1$ was
changed. Sometimes we will use $p_{10}$ instead of $p_0$ as the extra
parameter apart from $p_1,\dots p_{10-d}$.

In fact those two terms in (\ref{biltri}) are the only terms conserved by
T-dualities and only the relative ratio between them is undetermined.
However it is determined by S-dualities, which
exist for $d\leq 4$. For the moment, we ask the reader to
take this claim on faith.  Since the HE and HO moduli spaces
are the same on a torus, the same bilinear form can be viewed
in the $SO(32)$ language.  It takes the form (\ref{biltri}) in
$SO(32)$ variables as well.  

Let us note also another interesting invariance of 
(\ref{biltri}), which is useful for the $SO(32)$ case.
Let us express the parameters in the terms of the natural
parameters of the S-dual type~I theory
\eqn{sduone}{p_0=-q_0=-\ln(\gi), \qquad
p_i=q_i-\frac 12 q_0,\quad i=1,\dots 10-d}
where $q_i=\ln(R_i/ \li)$. We used $\gi=1/ \gh$ and $\lh=\gi^{1/2}\li$,
the latter expresses that the tension of the D1-brane and the heterotic
strings are equal. Substituing this into (\ref{biltri}) we get the same formula 
with $q$'s.
\eqn{qbiltri}{I=(-2q_0+\sum_{i=1}^{10-d}q_i)^2
+(d-2)\sum_{i=1}^{10-d}(q_i^2)}

\subsection{Moduli spaces and heterotic S-duality}

Let us recall a few well-known
facts about the moduli space of heterotic strings
toroidally compactified to $d$ dimensions. For $d>4$ the moduli space is
\eqn{modhet}{{\cal M}_d=\IR^+ \,\times\,
(SO(26-d,10-d,\IZ) \backslash SO(26-d,10-d,\IR) / SO(26-d,\IR)\times
SO(10-d,\IR)).}
The factor $\IR^+$ determines the coupling constant $\lh$. For $d=8$ the
second factor can be understood as the moduli space of elliptically fibered
K3's (with unit fiber volume), 
giving the duality with the F-theory. For $d=7$ the second factor
also
corresponds to the Einstein metrics on a K3 manifold with unit volume
which
expresses the duality with M-theory on K3. In this context, 
the factor $\IR^+$ can be
understood as the volume of the K3. Similarly for $d=5,6,7$ the
second factor describes conformal field theory of type~II string 
theories
on K3, the factor $\IR^+$ is related to the type~IIA coupling constant.

For $d=4$, i.e.\,compactification on $T^6$,
there is a new surprise. The field
strength
$H_{\kappa\lambda\mu}$ of the $B$-field can be Hodge-dualized to a 1-form
which
is the exterior derivative of a dual 0-form potential, the axion field.
The dilaton and axion are combined in the $S$-field which means that in
four noncompact dimensions,
 toroidally compactified heterotic strings exhibit
the $SL(2,\IZ)$ S-duality. 
\eqn{modhetfour}{{\cal M}_4=SL(2,\IZ)\backslash
SL(2,\IR)/SO(2,\IR) \,\times\,
(SO(22,6,\IZ) \backslash SO(22,6,\IR) / SO(22)\times
SO(6)).}

Let us find how our parameters $p_i$ transform under S-duality. 
The
S-duality
is a kind of electromagnetic duality. Therefore an electrically charged
state must be mapped to a magnetically charged state. The $U(1)$ symmetry
expressing rotations of one of the six toroidal coordinate is just one of
the 22 $U(1)$'s in the Cartan subalgebra of the full gauge group. It means
that the electrically charged states, the momentum modes in the given
direction of the six torus, must be mapped to the magnetically charged
objects which are the KK-monopoles.

The strings wrapped on the $T^6$ must be therefore mapped to the only
remaining point-like\footnote{Macroscopic strings (and higher-dimensional
objects) in $d=4$ have at least logarithmic IR divergence of the dilaton
and other fields and therefore their tension becomes infinite.}
 BPS objects available, i.e.\,to wrapped NS5-branes.
We know that NS5-branes are magnetically charged with respect to the
$B$-field so this action of the electromagnetic duality should not
surprise us. We find it convenient to combine this S-duality with
T-dualities on all six coordinates of the torus. The combined
symmetry $ST^6$ exchanges the point-like BPS objects in the
following way:

\begin{equation}
\begin{array}{|rcl|}
\hline \mbox{momentum modes} & \leftrightarrow & 
\mbox{wrapped NS5-branes}\\
\hline \mbox{wrapped strings} & \leftrightarrow & \mbox{KK-monopoles}\\
\hline 
\end{array}\label{tabulka}
\end{equation}

Of course, the distinguished direction inside the $T^6$ on both sides is
the same. The tension of the NS5-brane is equal to $1/(\gh^2\lh^6)$. Now consider 
the tension of the KK-monopole. In 11 dimensions,
a KK-monopole is reinterpreted as the D6-brane so its tension
must be
\eqn{tdsix}{T_{D6}=\frac{1}{g_{IIA}L_{IIA}^7}=\frac{R_{11}^2}{(\lpl)^9}}
where we have used $g_{IIA}=R_{11}^{3/2}\lpl^{-3/2}$
and $L_{IIA}=\lpl^{3/2}R_{11}^{-1/2}$ (from the tension of the fundamental
string).

The KK-monopole must always be a $(d-5)$-brane where $d$ is the
dimension of the spacetime. Since it is a gravitational object and the
dimensions along its worldvolume play no role, the tension must be always
of order $(R_1)^2$ in appropriate Planck units
where $R_1$ is the radius of the circle under whose
$U(1)$ the monopole is magnetically charged. Namely in the case of the
heterotic string in $d=4$, the KK-monopole must be another fivebrane whose
tension is equal to
\eqn{kkhet}{T_{KK5}=\frac{R_1^2}{(\lt)^8}=
\frac{{R_1}^2}{\gh^2\lh^8}}
where the denominators express the ten-dimensional Newton's constant.

Knowing this, we can find the transformation laws for $p$'s with respect
to the $ST^6$ symmetry. Here $V_6=R_1R_2R_3R_4R_5R_6$ denotes the volume
of the six-torus. Identifying the tensions in (\ref{tabulka}) we get

\eqn{vypocet}{\frac{1}{R'_1}=\frac{V_6}{\gh^2 R_1\lh^6},\qquad
\frac{R'_1}{(\lh')^2}= \frac{V_6 R_1}{\gh^2\lh^8}}

Dividing and multiplying these two equations we get respectively

\eqn{podil}{\frac{R'_1}{\lh'}=\frac{R_1}{\lh},\qquad
\frac{1}{\lh'}=\frac{V_6}{\gh^2\lh^7}.}

It means that the radii of the six-torus are fixed in string units
i.e.\,$p_1,\dots,p_6$ are fixed. Now it is straightforward to see that the
effective four-dimensional $SO(32)$ coupling constant $\gh^2 \lh^6/V_6$ is
inverted and the four-dimensional Newton's constant must remain unchanged.
The induced transformation on the $p$'s is
\eqn{indtr}{(p_0,p_1,\dots p_6, p_7,p_8 \dots)\to
(p_0+m,p_1,\dots p_6, p_7+m,p_8+m\dots)}
where $m=(p_1+p_2+p_3+p_4+p_5+p_6-2p_0)$ and the form (\ref{biltri}) 
can be checked to be constant.
It is also easy to see that such an invariance uniquely
determines the form up to an overall normalization i.e.\,it determines the
relative magnitude of two terms in (\ref{biltri}).

For $d=4$ this $ST^6$ symmetry can be expressed as $p_{10}\to -p_{10}$
with $p_1,\dots p_6$ fixed which gives the $\IZ_2$ subgroup of the
$SL(2,\IZ)$. For $d=3$ the transformation (\ref{indtr}) acts
as $p_7\leftrightarrow p_{10}$ so $p_{10}$ becomes one of eight parameters
that can be permuted with each other. It is a trivial consequence of the
more general fact that in three dimensions, the dilaton-axion field
unifies with the other moduli and the total space becomes
\cite{senthreed}
\eqn{modth}{{\cal M}_3=
SO(24,8,\IZ)\,\backslash\, SO(24,8,\IR)\, /\, SO(24,\IR)\times SO(8,\IR).}

We have thus repaid our debt to the indulgent reader, and 
verified that the bilinear form (\ref{biltri}) is
indeed invariant under the dualities of the heterotic moduli
space for $d \geq 3$.  For $d=2$ the bilinear form is degenerate
and is the Cartan form of the affine algebra $\hat{o}(8,24)$
studied by \cite{sentwod}. For $d=1$ it is the Cartan form
of $DE_{18}$ \cite{ori}.  The consequences of this for the
structure of the extremes of moduli space are nearly identical to
those of \cite{bfm}.  The major difference is our relative lack
of understanding of the safe domain.  We believe that this
is a consequence of the existence of 
regimes like F-theory or 11D SUGRA  on a large smooth K3 
with isolated singularities, where much of the physics is
accessible but there is no systematic expansion of all
scattering amplitudes.   In the next section we make some
remarks about different extreme regions of the restricted moduli
space that preserves the full $SO(32)$ symmetry.

\section{Covering the $SO(32)$ moduli space}

\subsection{Heterotic strings, type~I, type~IA and $d\geq 9$}

One new feature of heterotic moduli spaces is the apparent possibility
of having asymptotic domains with enhanced gauge symmetry.  For example,
if we consider the description of heterotic string theory on a torus
from the usual weak coupling point of view, there are domains with 
asymptotically large heterotic radii and weak coupling, where the
the full nonabelian rank $16$ Lie groups are restored.  All other 
parameters are held fixed at what appears from the weak coupling
point of view to be \lq\lq generic\rq\rq values.  This includes Wilson 
lines.  In the large volume limit, local physics is not
sensitive to the Wilson line symmetry breaking. 

Now, consider the limit
described by weakly coupled Type IA string theory on a large orbifold.  
In this limit, the
theory consists of D-branes and orientifolds, placed along a line
interval.   There is no way to restore the $E_8\times E_8$ symmetry
in this regime.  Thus, even the safe domain of asymptotic moduli space 
appears to be divided into regimes in which different nonabelian
symmetries are restored.  Apart from sets of measure zero ({\it e.g.}
partial decompactifications) we either have one of the full rank $16$
nonabelian groups, or no nonabelian symmetry at all.  The example of
F-theory tells us that the abelian portion of asymptotic moduli space 
has regions without a systematic semiclassical expansion.  

In a similar manner, consider 
the moduli space of the $E_8\times E_8$ heterotic strings
on rectilinear tori.  We have only two semiclassical descriptions with
manifest $E_8\times E_8$ symmetry, namely HE strings and the Ho\hacek
rava-Witten (HW) domain walls. Already for $d=9$ (and any $d<9$) we would
find limits that are described neither by HE nor by HW. For example, 
consider a limit of M-theory on a cylinder with very large $\gh$ but the radius
of the circle, $R$, in the domain $L_P \gg R \gg \lh^2/ \lpl $, and
unbroken $E_8\times E_8$.  We do not know how to describe this limit with
any known semiclassical expansion.  We will find that we can get a more 
systematic description of asymptotic domains in the HO case, and will
restrict attention to that regime for the rest of this paper.

\FIGURE[l]{\epsfig{file=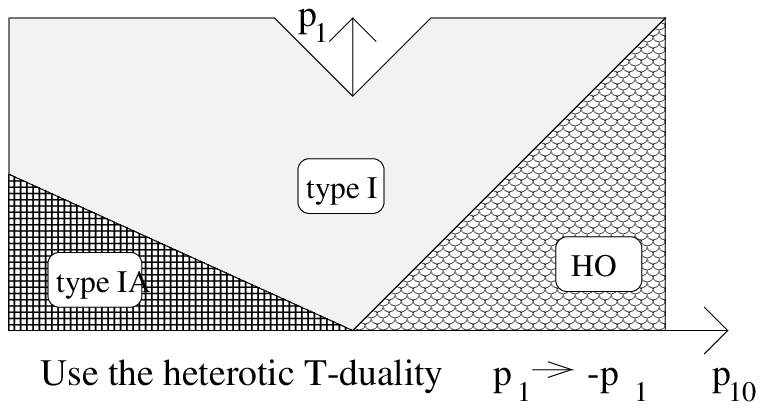}\caption{The limits
in $d=9$.}\label{ninefigure}}

For $d=10$ there are only two limits.  $p_0<0$ gives the
heterotic strings and $p_0>0$ is the type I theory. However already for
$d=9$ we have a more interesting picture analogous to the figure 1
in \cite{bfm}. Let us make a counterclockwise trip around the figure.
We start at a HO point with $p_1=0$ which is a weakly coupled heterotic
string
theory with radii of order $\lh$ (therefore it is adjacent to its T-dual
region). When we go around the circle, the radius and also the coupling
increases and we reach the line $p_0\equiv (p_1-p_{10})/2=0$ where we
must switch to the type I description. Then the radius decreases again so
that we must perform a T-duality and switch to the type~IA
description. This happens for $p_1-(p_0/2)=(3p_1+p_{10})/4=0$; we had to
convert $R_1$ to the units of $\li=\gh^{1/2}\lh$. Then we go on and the
coupling $g_{IA}$ and/or the size of the line interval increases. The most
interesting is the final boundary given by $p_1=0$ which guarantees that
each of the point of the $p$-space is covered precisely by one limit.

\vspace{3mm}

We can show that $p_1>0$ is precisely the condition that the dilaton in
the type~IA theory is not divergent. Rou\-g\-h\-ly speaking, in units of
$\li=L_{IA}$ the ``gravitational potential'' is linear in $x_1$ and
proportional to $g_{IA}^2 / g_{IA}$. Here $g_{IA}^2$ comes from
the gravitational constant and $1/g_{IA}$ comes from the tension of the
D8-branes. Therefore we require not only $g_{IA}<1$ but also 
$g_{IA}<\li / R_{line\,interval}$. Performing the T-duality
$\li / R_{line\,interval}=R_{circle} / \li$ and converting to $\lh$
the condition becomes precisely $R_{circle}>\lh$.

In all the text we adopt (and slightly modify) the standard definition
\cite{bfm} for an
asymptotic description to be viable: dimensionless coupling
constants should be smaller than one, but in cases without translational
invariance, the dilaton should not diverge anywhere, and the
sizes of the effective geometry should be greater than the appropriate
typical scale (the string length for string theories or the Planck length
for M-theory). It is important to realize that in the asymptotic
regions we can distinguish between e.g. type~I and type~IA 
because their physics is different. We cannot distinguish between them
in case the T-dualized circle is of order $\li$ but such vacua are of
measure zero in our investigation and form various boundaries in the
parameter space.  This is the analog of the distinction we made between
the IIA and IIB asymptotic toroidal moduli spaces in \cite{bfm} 

\subsection{Type IA${}^2$ and $d=8$}

In $d=8$ we will have to use a new desciption to cover the parameter
space, namely the double T-dual of type I which we call type~IA${}^2$.
Generally, type~IA${}^k$ contains 16 D-$(9-k)$-branes, their images and
$2^k$ orientifold
$(9-k)$-planes. We find it also useful to perform heterotic T-dualities to
make $p_i$ positive for $i=1,\dots, 10-d$ and sort $p$'s so that our
interest is (without a loss of generality) only in configurations with
\eqn{sort}{0\leq p_1\leq p_2\leq \dots \leq p_{10-d}}
We need positive $p$'s for the heterotic description to be valid but such
a transformation can only improve the situation also for type~I and its
T-dual
descriptions since when we turn $p$'s from negative to positive values,
$\gh$ increases and therefore $\gi$ decreases. For type~I we also need
large radii. For its T-duals we need a very small string coupling and if
we make a T-duality to convert $R>\li$ into $R< \li$, the coupling 
$g_{IA}$ still decreases; therefore it is good to have as large radii in
the type~I limit as possible.

\vspace{3mm}

In $d=8$ our parameters are $p_0,p_1,p_2$ or $p_{10},p_1,p_2$ where
$p_{10}=-2p_0+p_1+p_2$ and we will assume $0< p_1< p_2$ as we have
explained (sets of measure zero such as the boundaries between regions
will be neglected). If $p_0<0$, the HO description is good. Otherwise
$p_0>0$. If furthermore $2p_1-p_0>0$ (and therefore also $2p_2-p_0>0$),
the radii are large in the type~I units and we can use the (weakly
coupled) type~I description. Otherwise $2p_1-p_0<0$. If furthermore
$2p_2-p_0>0$, we can use type~IA strings. Otherwise $2p_2-p_0<0$ and the
type~IA${}^2$ description is valid. Therefore we cover all the parameter
space. Note that the F-theory on K3 did not appear here.  In asymptotic
moduli space, the F-theory regime generically has no enhanced nonabelian
symmetries. 

\vspace{3mm}

In describing the boundaries of the moduli space, we used the relations 
$\lh=\gi^{1/2}\li$, $\gh=1/\gi$. 
The condition for the dilaton not to diverge is still $p_1>0$
for any type~IA${}^k$ description. The longest direction of the $T^k/
\IZ_2$ of this theory is still the most dangerous for the dilaton
divergence and is not affected by the T-dualities on the shorter
directions of the $T^k/ \IZ_2$ orientifold. For $d=9$ (and fortunately
also for $d=8$)
the finiteness of the dilaton field automatically implied that
$g_{IA^k} < 1$. However this is not true for general $d$. After a short
chase through a sequence of S and T-dualities we find that the condition
$g_{IA^k}<1$ 
can be written as
\eqn{karkulka}{(k-2)p_0 -2\sum_{i=1}^k p_i<0.}
We used the trivial requirement that the T-dualities must be performed on
the shortest radii (if $R_j< \li$, also $R_{j-1} < \li$ and therefore it
must be also T-dualized). Note that for $k=1$ the relation is
$-p_0-2p_1<0$ which is a trivial consequence of $p_1>0$ and $p_0>0$.
Also for $k=2$ we get a trivial condition $-2(p_1+p_2)<0$. However for
$k>2$ this condition starts to be nontrivial. This is neccessary for
consistency: otherwise IA${}^k$ theories would be sufficient to cover
the whole asymptotic moduli space, and because
of S-dualities we would cover the space several times. It would be also
surprising not to encounter regimes described by large K3 geometries.

\subsection{Type IA${}^3$, M-theory on K3 and $d=7$}

This happens already for $d=7$ where the type~IA${}^3$
description must be added. The reasoning starts in the same way:
for $p_0<0$ HO, for $2p_1-p_0>0$ type~I, for $2p_2-p_0>0$ type~IA, 
for $2p_3-p_0>0$ type~IA${}^2$.

However, when we have $2p_3-p_0<0$ we cannot deduce that the conditions for
type~IA${}^3$ are obeyed because also (\ref{karkulka}) must be imposed:
\eqn{kremilek}{p_0-2(p_1+p_2+p_3)<0}
It is easy to see that this condition is the weakest one i.e.\,that it is
implied by any of the conditions
$p_0<0$, $2p_1-p_0>0$, $2p_2-p_0>0$ or $2p_3-p_0>0$.
Therefore the region that we have not covered yet is given by the opposite
equation
\eqn{vochomurka}{2p_0-4(p_1+p_2+p_3)=-p_{10}-3(p_1+p_2+p_3)>0}
The natural hypothesis is that this part of the asymptotic parameter space
is the limit where we can use the description of M-theory on a K3
manifold. However things are not so easy: the condition that $V_{K3}>
(\lpl)^4$ gives just $p_{10}<0$ which is a weaker requirement
than (\ref{vochomurka}).

The K3 manifold has a $D_{16}$ singularity but this is not the real source
of the troubles. A more serious issue is that the various typical sizes of
such a K3 are very different and we should require that each of them is
greater than $\lpl$ (which means that the shortest one is). In an analogous
situation with $T^4$ instead of K3 the condition $V_{T^4}>\lpl^4$ would be
also insufficient: all the radii of the four-torus must be greater
than $\lpl$.

Now we would like to argue that the region defined by (\ref{vochomurka})
with our gauge $0<p_1<p_2<p_3$ can indeed be described by the 11D SUGRA 
on K3, except near the $D_{16}$ singularity.  Therefore, all of the
asymptotic moduli space is covered by regions which have a reasonable
semiclassical description.

While the fourth root of the volume of K3 equals
\eqn{arabela}{\frac{V_{K3}^{1/4}}{\lpl}=\frac{\gh^{1/3}\lh^{1/2}}{V_3^{1/6}}
=\exp\left(p_0/3-(p_1+p_2+p_3)/6\right)=\exp(-p_{10}/6),}
the minimal typical distance in K3 must be corrected to agree with
(\ref{vochomurka}). We must correct it only by a factor depending on the
three radii in heterotic units (because only those are the parameters in
the moduli space of metric on the K3) so the distance equals (confirming
(\ref{vochomurka}))
\eqn{rumburak}{\frac{L_{min.K3}}{\lpl}=\exp\left(-p_{10}/6-(p_1+p_2+p_3)/2
\right).}
Evidence that (\ref{rumburak}) is really correct and
thus that we understand the limits for $d=7$ is the following.
We
must first realize that 16 independent two-cycles are shrunk to zero size
because of the $D_{16}$ singularity present in the K3 manifold. This
singularity implies a lack of understanding of the physics in a vicinity
of this point but it does not prevent us from describing the physics in the
rest of K3 by 11D SUGRA. So we allow
the 16 two-cycles to shrink. The
remaining 6 two-cycles generate a space of signature 3+3 in the cohomology
lattice: the intersection
numbers are identical to the second cohomology of $T^4$. We can compute
the areas of those 6 two-cycles because the M2-brane wrapped on the
6-cycles are dual to the wrapped heterotic strings and their momentum
modes. Now let us imagine that the geometry of the two-cycles of K3 can be
replaced by the 6 two-cycles of a $T^4$ which have the same intersection
number.

\vspace{3mm}

It means that the areas can be written as
$a_1a_2,a_1a_3,a_1a_4$, $a_2a_3,a_2a_4,a_3a_4$ where $a_1,a_2,a_3,a_4$
are the radii of the four-torus and correspond to some typical distances
of the K3. If we order the $a$'s so that $a_1<a_2<a_3<a_4$, we see that the
smallest of the six areas is $a_1a_2$ (the largest two-cycle is the dual
$a_3a_4$) and similarly the second smallest area is $a_1a_3$ (the second
largest two-cycle is the dual $a_2a_4$). On the heterotic side we have
radii $\lh<R_1<R_2<R_3$ (thus also $\lh^2/R_3< \lh^2/R_2< \lh^2/R_1<\lh$)
and therefore the correspondence between the
membranes and the wrapping and momentum modes of heterotic strings tells
us that
\eqn{budulinek}{\frac{a_1a_2}{\lpl^3}=\frac{1}{R_3},
\quad \frac{a_3a_4}{\lpl^3}=\frac{R_3}{\lh^2},
\qquad
\frac{a_1a_3}{\lpl^3}=\frac{1}{R_2},
\quad \frac{a_2a_4}{\lpl^3}=\frac{R_2}{\lh^2}.}
As a check, note that $V_{K3}=a_1a_2a_3a_4$ gives us $\lpl^6/ \lh^2$ as
expected (since heterotic strings are M5-branes wrapped on $K3$).
We will also assume that
\eqn{liska}{\frac{a_1a_4}{\lpl^3}=\frac{1}{R_1}, 
\quad \frac{a_2a_3}{\lpl^3}=\frac{R_1}{\lh^2}.}
Now we can calculate the smallest typical distance on the K3.
\eqn{jezinka}{a_1=\sqrt{\frac{a_1a_2\cdot a_1a_3}{a_2a_3}}=
\frac{\lpl^{3/2}\lh}{\sqrt{R_1R_2R_3}}}
which can be seen to coincide with (\ref{rumburak}). There is a subtlety
that we should mention. It is not completely clear whether $a_1a_4<a_2a_3$
as we assumed in (\ref{liska}). The opposite possibility is obtained by
exchanging $a_1a_4$ and $a_2a_3$ in (\ref{liska}) and leads to $a_1$ 
greater than (\ref{jezinka}) which would imply an overlap with the other
regions. Therefore we believe that the calculation in
(\ref{liska}) and (\ref{jezinka}) is the correct way to find the condition
for the
K3 manifold to be large enough for the 11-dimensional supergravity (as a
limit of M-theory) to be a good description.

\subsection{Type IA${}^{4,5}$, type~IIA/B on K3 and $d=6,5$}

Before we will study new phenomena in lower dimensions, it is useful to
note that in
any dimension we add new descriptions of the physics.
The last added limit always corresponds to the ``true'' S-dual of the
original heterotic string theory -- defined by keeping the radii fixed in
the heterotic string units (i.e.\,also keeping the shape of the K3
geometry) and sending the coupling to infinity -- because this last limit
always contains the direction with $p_0$ large and positive (or $p_{10}$
large and negative) and other $p_i$'s much smaller.

\begin{itemize}
\item In 10 dimensions, the true S-dual of
heterotic strings is the type~I theory.
\item In 9 dimensions it is type~IA.
\item In 8 dimensions type~IA${}^2$.
\item In 7 dimensions we get M-theory on K3.
\item In 6 dimensions type IIA strings on K3.
\item In 5 dimensions type IIB strings on K3$\times S^1$
where the circle decompactifies as the coupling goes to infinity. The
limit is therefore a six-dimensional theory.
\item In 4 dimensions we observe a mirror copy of the region $p_{10}<0$
to arise for $p_{10}>0$. The strong coupling limit is the heterotic
string itself.
\item In 3 dimensions the dilaton-axion is already unified with the other
moduli so it becomes clear that we studied an overly specialized
 direction in the
examples above. Nevertheless the same claim as in $d=4$ can be made.
\item In 2 dimensions only positive values of $p_{10}$ are possible
therefore the strong coupling limit does not exist in the safe domain 
of moduli space. 
\item In 1 dimension the Lorentzian structure of the parameter space
emerges. Only the future light cone corresponds to semiclassical physics
which is reasonably well understood.
The strong coupling limit defined above would lie inside the unphysical
past light cone.
\end{itemize}

Now let us return to the discussion of how to separate the parameter
space into regions where different semiclassical 
descriptions are valid. We may repeat
the same inequalities as in $d=7$ to define the limits HO, I, IA, IA${}^2$,
IA${}^3$. But for M-theory on K3 we must add one more condition
to the constraint (\ref{vochomurka}): a new circle has been added and
its size
should be also greater than $\lpl$. For the new limit of the type~IIA
strings on K3 we encounter similar problems as in the case of the
M-theory on K3. Furthermore if we use the definition (\ref{rumburak})
and postulate this shortest distance to be greater than the type~IIA
string length, we do not seem to get a consistent picture covering the whole
moduli space. Similarly for $d=5$, there appear two new asymptotic
descriptions, namely type~IA${}^5$ theory and type~IIB strings on
$K3\times S^1$. It is clear that the condition $g_{IA^5}<1$ means
part of the parameter space is not understood and another description,
most probably type~IIB strings on $K3\times S^1$, must be used.
Unfortunately at this moment we are not able to show that the condition
for the IIB theory on K3 to be valid is complementary to the condition
$g_{IA^5}<1$. A straightforward application of (\ref{jezinka}) already for
the type~IIA theory on a K3 gives us a different inequality. Our lack
of understanding of the limits for $d<7$ might be solved by employing
a correct T-duality of the type~IIA on K3 but we do not have a complete
and consistent picture at this time.

\subsection{Type IA${}^6$ and S-duality in $d=4$}

Let us turn to the questions that we understand better.
As we have already said, in $d=4$ we see the $\IZ_2$ subgroup of the
$SL(2,\IZ)$ S-duality which acts as $p_{10}\to -p_{10}$ and
$p_1,\dots,p_6$ fixed in our formalism. This reflection divides
the $p$-space
to subregions $p_{10}>0$ and $p_{10}<0$ which will be exchanged
by the S-duality. This implies that a new description should require
$p_{10}>0$. Fortunately this is precisely what happens: in $d=4$ we have
one new limit, namely the type~IA${}^6$ strings and the condition
(\ref{karkulka}) for $g_{IA^6}<1$ gives
\eqn{myslivec}{4p_0-2\sum_{i=1}^6 p_i =-2p_{10}<0}
or $p_{10}>0$.

\vspace{3mm}


In the case of $d=3$ we find also a fundamental domain that is copied
several times by S-dualities. This fundamental region is again bounded by
the condition $g_{gauge}^{eff.4-dim}<1$ which is the same like
$g_{IA^6}<1$ and the internal structure has been partly described:
the fundamental region is divided into several subregions HO, type~I,
type~IA${}^k$, M/K3, IIA/K3, IIB/K3. As we have said, we do not understand
the limits with a K3 geometry well enough to separate the fundamental
region into the subregions enumerated above. We are not even sure whether
those limits are sufficient to cover the whole parameter space. In the case of
$E_8\times E_8$ theory, we are pretty sure that there are some limits that
we do not understand already for $d=9$ and similar claim can be true
in the case of the $SO(32)$ vacua for $d<7$.  We understand much
better how the entire parameter space can be divided into the copies of the
fundamental region and we want to concentrate on this question.

The inequality $g_{gauge}^{eff.4-dim}<1$ should hold
independently of which of the six radii are chosen to be the radii
of the six-torus. In other words, it must hold for the smallest radii and
the condition is again (\ref{myslivec}) which can be for $d=3$ reexpressed
as $p_{7}<p_{10}$.

So the ``last'' limit at the boundary of the fundamental region is again
type~IA${}^6$ and not type~IA${}^7$, for instance. It is easy to show that
the condition $g_{IA^6}<1$ is implied by any of the conditions for the
other limits so this condition is the weakest of all: all the regions
are inside $g_{IA^6}<1$.

This should not be surprising, since
$g_{gauge}^{eff.4-dim}=(g_{IA^6})^{1/2}
=g_{IA^6}^{open}$; the heterotic S-duality in this type~IA${}^6$ limit can
be identified with the S-duality of the effective low-energy description
of the D3-branes of the type~IA${}^6$ theory.
As we have already said, this inequality reads for $d=3$
\eqn{cipisek}{2p_0-\sum_{i=1}^6 p_i =-p_{10}+p_7<0}
or $p_{10}>p_7$. We know that precisely in $d=3$ the S-duality (more
precisely the $ST^6$ transformation) acts as the permutation of $p_7$
and $p_{10}$. Therefore it is not hard to see what to do if we want to
reach the
fundamental domain: we change all signs to pluses by T-dualities and sort
all {\it eight} numbers $p_1,\dots p_7; p_{10}$ in the ascending order.
The inequality (\ref{cipisek}) will be then satisfied. The condition
$g_{gauge}^{eff.4-dim}<1$ or (\ref{myslivec}) will  define the
fundamental region also for the case of one or two dimensions.

\subsection{The infinite groups in $d\leq 2$}

In the dimensions $d>2$ the bilinear form is positive
 definite and the
group of dualities conserves the lattice $\IZ^{11-d}$ in 
the $p$-space.
Therefore the groups are finite. However for $d=2$ (and {\it a fortiori}
for $d=1$ because the $d=2$ group is isomorphic to 
a subgroup of the $d=1$ group) the
group becomes infinite. In this dimension $p_{10}$ is unchanged
by T-dualities and S-dualities. The
regions with $p_{10}\leq 0$ again correspond to 
mysterious regions where the holographic principle appears to be
violated, as in \cite{bfm}. Thus we may assume
that $p_{10}=1$; the overall normalization does not matter.

Start for instance with $p_{10}=1$ and
\eqn{rumcajs}{(p_1,p_2,\dots p_8)=(0,0,0,0,0,0,0,0)}
and perform the S-duality ($ST^6$ from the formula (\ref{indtr})) with
$p_7$ and $p_8$ understood as the large
dimensions (and $p_1\dots p_6$ as the 6-torus). This transformation
maps $p_7\mapsto p_{10}-p_8$ and $p_8\mapsto p_{10}-p_7$. So if we repeat
$ST^6$ on $p_7,p_8$, T-duality of $p_7,p_8$, $ST^6$, $T^2$ and so on,
$p_{1}\dots p_6$ will be still zero and the values of $p_7,p_8$ are
\eqn{rakosnicek}{(p_7,p_8)=(1,1)\to(-1,-1)\to(2,2)\to(-2,-2)\to(3,3)
\to\dots}
and thus grow linearly to infinity, proving the infinite order of the
group. The equation for $g_{IA^6}<1$ now gives
\eqn{bobek}{2p_0-\sum_{i=1}^6 p_i =-p_{10}+p_7+p_8<0}
or $p_{10}>p_7+p_8$. Now it is clear how to get to such a fundamental
region with (\ref{bobek}) and $0<p_1<\dots p_8$. We repeat the $ST^6$
transformation with the two largest radii ($p_7,p_8$)
as the large coordinates. After each step we turn the signs to $+$ by
T-dualities and order $p_1<\dots< p_8$ by permutations of radii.
A bilinear quantity decreases assuming $p_{10}>0$ and $p_{10}<p_7+p_8$
much like in \cite{bfm}, the case $k=9$ ($d=2$):
\eqn{pokuston}{C_{d=2}=\sum_{i=1}^8 (p_i)^2\to
\sum_{i=1}^8 (p_i)^2+2p_{10}(p_{10}-(p_7+p_8))}
In the same way as in \cite{bfm},
starting with a rational approximation of a vector $\vec p$, 
the quantity $C_{d=2}$
cannot decrease indefinitely and therefore finally we must get to a point
with $p_{10}>p_7+p_8$. 

\vspace{3mm}

In the case $d=1$ the bilinear form has a Minkowski signature. The
fundamental region is now limited by 
\eqn{hurvinek}{2p_0-\sum_{i=1}^6 p_i =-p_{10}+p_7+p_8+p_9<0}
and it is easy to see that under the $ST^6$ transformation on radii
$p_1\dots p_6$, $p_{10}$ transforms as
\eqn{spejbl}{p_{10}\to 2p_{10}-(p_7+p_8+p_9).}
Since the $ST^6$ transformation is a reflection of a spatial coordinate in
all cases, it keeps us inside the future light cone if we start there.
Furthermore, after each step we make such T-dualities and permutations
to ensure $0<p_1<\dots p_9$.

If the initial $p_{10}$ is greater than $[(p_1)^2+\dots+(p_9)^2]^{1/2}$
(and therefore positive), it remains positive and assuming $p_{10}<p_7+p_8
+p_9$, it decreases according to (\ref{spejbl}). But it cannot decrease
indefinitely (if we approximate $p$'s by rational numbers or integers
after a scale transformation). So at some point the assumption
$p_{10}<p_7+p_8+p_9$ must break down and we reach 
the conclusion that fundamental domain is characterized by
$p_{10}>p_7+p_8+p_9$.

\subsection{The lattices}

In the maximally supersymmetric case \cite{bfm}, we encountered
exceptional algebras and their corresponding lattices. We were able to see some
properties of the Weyl group of the exceptional algebra $E_{10}$ and
define its fundamental domain in the Cartan subalgebra. In the present
case with 16 supersymmetries, the structure of lattices for $d>2$ is not
as 
rich. The dualities always map integer vectors $p_i$ onto integer
vectors.

For $d>4$, there are no S-dualities and our T-dualities know about the
group $O(26-d,10-d,\IZ)$. For $d=4$ our group contains an extra $\IZ_2$
factor from the single S-duality. For $d=3$ they unify to a larger group
$O(8,24,\IZ)$. We have seen the semidirect product of $(\IZ_2)^8$ and
$S_8$ related to its Weyl group in our formalism.
For $d=2$ the equations of motion exhibit a larger
affine $\hat o(8,24)$ algebra whose discrete duality group
has been studied in \cite{sentwod}.

In $d=1$ our bilinear form has Minkowski signature. The S-duality can be
interpreted as a reflection with respect to the vector
\eqn{refsl}{(p_1,p_2,\dots, p_9,p_{10})=(0,0,0,0,0,0,-1,-1,-1,+1).}
This is a spatial vector with length-squared equal to minus two
(the form (\ref{biltri}) has a time-like signature). As we have seen,
such reflections generate together with T-dualities an infinite group
which is an evidence for an underlying hyperbolic algebra 
analogous to $E_{10}$.
Indeed, Ganor \cite{ori} has argued that the $DE_{18}$ ``hyperbolic''
algebra underlies the nonperturbative duality group of maximally
compactified heterotic string theory.  The Cartan algebra of this
Dynkin diagram unifies the asymptotic directions which we have studied
with compact internal symmetry directions.  Its Cartan metric has one
negative signature direction.

\section{Conclusions}

The parallel structure of the moduli spaces with 32 and 16 SUSYs
gives us reassurance that the features uncovered in \cite{bfm}
are general properties of M-theory.  It would be interesting
to extend these arguments to moduli spaces with less SUSY.
Unfortunately, we know of no algebraic characterization of the
moduli space of M-theory on a Calabi Yau threefold.  Furthermore, 
this moduli space is no longer an orbifold.  It is stratified,
with moduli spaces of different dimensions connecting to each 
other via extremal transitions.  Furthermore, in general the
metric on moduli space is no longer protected by 
nonrenormalization theorems, and we are far from a 
characterization of all the extreme regions.  For the case
of four SUSYs the situation is even worse, for most of what
we usually think of as the moduli space actually has
a superpotential on it, which generically is of order the
fundamental scale of the theory.
\footnote{Apart from certain extreme regions,
where the superpotential asymptotes to zero,
the only known loci on which it vanishes 
are rather low dimensional
subspaces of the classical moduli space, \cite{bdw}.}

There are thus many hurdles to be jumped before we can claim
that the concepts discussed here and in \cite{bfm} 
have a practical application to realistic cosmologies.

\vspace{15mm}

\acknowledgments

We are grateful to Ori Ganor
for valuable discussions. This work 
was supported in part by the DOE under grant
number DE-FG02-96ER40559. 

\newpage



\end{document}